\documentclass{aastex}

\usepackage{epsfig}
\usepackage{amssymb}
\usepackage{gensymb}
\usepackage{color}
\definecolor{red}{rgb}{0.7,0,0}
\definecolor{blue}{rgb}{0,0,0.7}
\definecolor{purple}{rgb}{0.7,0,0.7}

\def\refee#1{{{#1}}}
\def\correc#1{{{#1}}}

\def\chisq{$\chi^2_\nu$} 
\def\rxte{{\em {RXTE}}}

\def\XTE{XTE~J1550$-$564}

\shorttitle{Elusive HFQPOs in XTE J1550$-$564}
\shortauthors{Varniere \& Rodriguez}
 
\begin{document}

\title{Looking for the elusive $3$:$2$ ratio of high-frequency quasi-periodic oscillations in the microquasar XTE J1550$-$564}

\author{Peggy Varniere}
\affil{APC, AstroParticule et Cosmologie, Universit\'e Paris Diderot, CNRS/IN2P3, 
CEA/Irfu, Observatoire de Paris, Sorbonne Paris Cit\'e, 10, rue Alice Domon et LŽonie Duquet, 75205 Paris Cedex 13, France }
\affil{AIM, CEA, CNRS, UniversitŽ Paris-Saclay, Universit\'e Paris Diderot, Sorbonne Paris Cit\', F-91191 Gif-sur-Yvette, France.}

\and 

\author{J\'er\^ome Rodriguez} 
\affil{AIM, CEA, CNRS, UniversitŽ Paris-Saclay, Universit\'e Paris Diderot, Sorbonne Paris Cit\', F-91191 Gif-sur-Yvette, France.}

\email{varniere@apc.univ-paris7.fr, jrodriguez@cea.fr}

 
\date{Received  /
Accepted }

\begin{abstract}
 Using the two main XTE J1550$-$564 outbursts  (1998-99 and 2000) we gathered about 30 observations with confirmed 
 detections of  HFQPOs. While this is a small sample it is enough to start looking at the generic properties of these 
 oscillations, especially focusing on their frequencies and their potential harmonic relationship. This then will provide us with   
 a list of constraints, necessary for any attempt of modelling their origin.  
We defined five groups based on their similarities in the Fourier domain, namely the continuum of their power density 
spectra (PDS) and the HFQPO frequencies. 
We then combined the individual PDSs of each family to obtain a PDS with higher statistics  to  search for other potential, 
previously undetected, weaker peaks.
While we have two $3\sigma$ potential detections of a  pair of HFQPOs in our combined PDSs, none of them shows HFQPOs with frequencies in a previously claimed 3:2 ratio.
Using the results presented here we propose an updated list of requirements for any model trying to explain the HFQPOs in microquasars.
\end{abstract}


\section{Introduction}
 Since their first detection there have been a long string of efforts to understand the source of the variability observed in microquasars 
 but no model has yet  gained wide acceptance. {This is especially true for the origin of the rather elusive} High-Frequency Quasi-Periodic 
 Oscillation (HFQPO) {in systems containing (or thought to contain) a black hole (BH) as compact object}. 
 HFQPOs appear as narrow peak(s) in the X-ray power-density spectra (PDS) of BH binaries.  {They have been detected in eight different 
 BH sources \citep[GRO J1655$-$40, GRS 1915+105, XTE J1550$-$564, H1743$-$322, 4U 1630$-$47, XTE J1650$-$500, 
 XTE J1859+226 IGR J17091$-$3624, e.g. ][and references therein]{Remillard06,altamirano12,belloni12}, ranging from 
as low as 27~Hz in GRS 1915+105  \citep{belloni2001}  up to a  few hundred Hz \citep[a maximum of 450 Hz was seen in GRO J1655$-$40][]
 {Remillard06}}.  HFQPOs are  particularly interesting as their frequencies typically lie in the frequency range of the Keplerian 
 frequency of the last stable orbit around the central BH. They can therefore be seen as a window 
 to the innermost region of accretion where strong gravity is expected to play an important role\footnote{Relativistic effects, and their influences on the properties of low frequency QPOs, are, for example, discussed within the context of the accretion-ejection instability in \citet{VTR12}.}. 
 This is one of the reasons 
 why HFQPOs {have stimulated much more interest than their low frequency counterparts (LFQPO)} even though  they are 
 much weaker and rarer. \\
\indent Somes sources, and especially XTE J$1550$-$564$, have {exhibited HFQPOs} with enough regularity to start studying their properties 
with respect to the source's spectral states and also search for groups (or families) of similar PDSs that can then be combined in order to
improve the statistics and probe fainter components. 
Here we aim at presenting a comprehensive study of the occurrences of HFQPOs in the outburst of XTE J$1550$-$564$. 
While most of the data are  taken from published papers \citep{remillard02b, miller01_1550, Sobczak00,rodrigue03_1550,belloni12}, 
we reanalyzed and combined some observations together to try and 
obtain better statistical  significances on the presence and obtained parameters of already known XTE J$1550$-$564$'s HFQPOs. 
Using the combined PDSs we then study the possible simultaneous presence of multiple high frequency peaks (or pairs of HFQPOs), 
and if present investigate the ratio of their frequencies.  
Using five groups of observations we can then see how the presence/absence of an additional  
peak improves, and increases the constraints already known on models of HFQPOs.
We then conclude this paper with a short list of constraints that any model trying to explain HFQPOs 
should respect. 
    
\section{Looking for the elusive $2$:$3$ relationship between the HFQPOs frequencies}

	Many authors have pointed out a ratio 3:2 between different frequencies of HFQPOs in various sources such as XTE J$1550$-$564$ \citep{miller01_1550} and GRO J1655$-$40 \citep{remillard02b}. 
	Such pairs are, however, not always detected, and  in some reported cases they can show different frequency 
	ratios than the 3:2 e.g. \citet{Strohmayer01b} and \citet{belloni_alt2013} in the case of  GRS 1915+105.  
	Since the predicted systematic presence of a specific ratio, such as the 3:2  would be a strong constraint 
	on theoretical models we decided to explore this aspect in more detail.
	In \XTE\  HFQPOs have been detected at various frequencies (from $\sim$100~Hz to $\sim$280~Hz) both 
	during the major 1998--1999, and fainter 2000 outbursts \citep{remillard02b, miller01_1550}. \citet{belloni12}
	analyse the properties of HFQPOs of various BH sources. In addition to HFQPOs reported earlier (for which they 
	obtain compatible frequencies), they also report new occurrences of HFQPOs when analysing only the hard ($\gtrsim 6$ keV) 
	band light curves in four additional observations. 
	Pairs of HFQPOs have, however, not been detected  in any of the individual observations 
	considered in these studies, but they were reported from stacked observations, as we describe below.
	
\begin{figure}[htbp]
\centering
\begin{tabular}{cc}
\multicolumn{2}{c}{\resizebox{8.5cm}{!}{\includegraphics{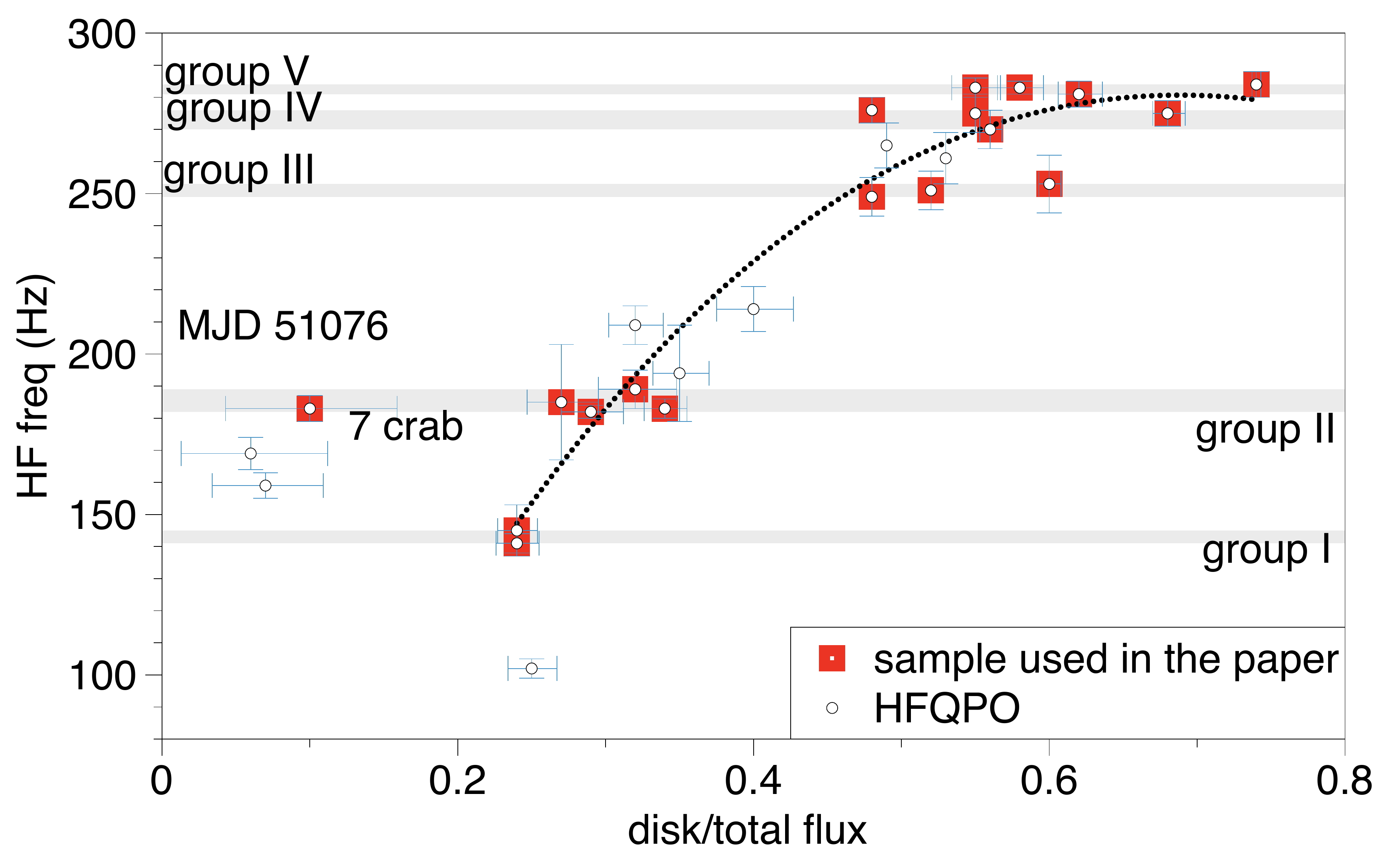}}}\\
\resizebox{4.cm}{!}{\includegraphics{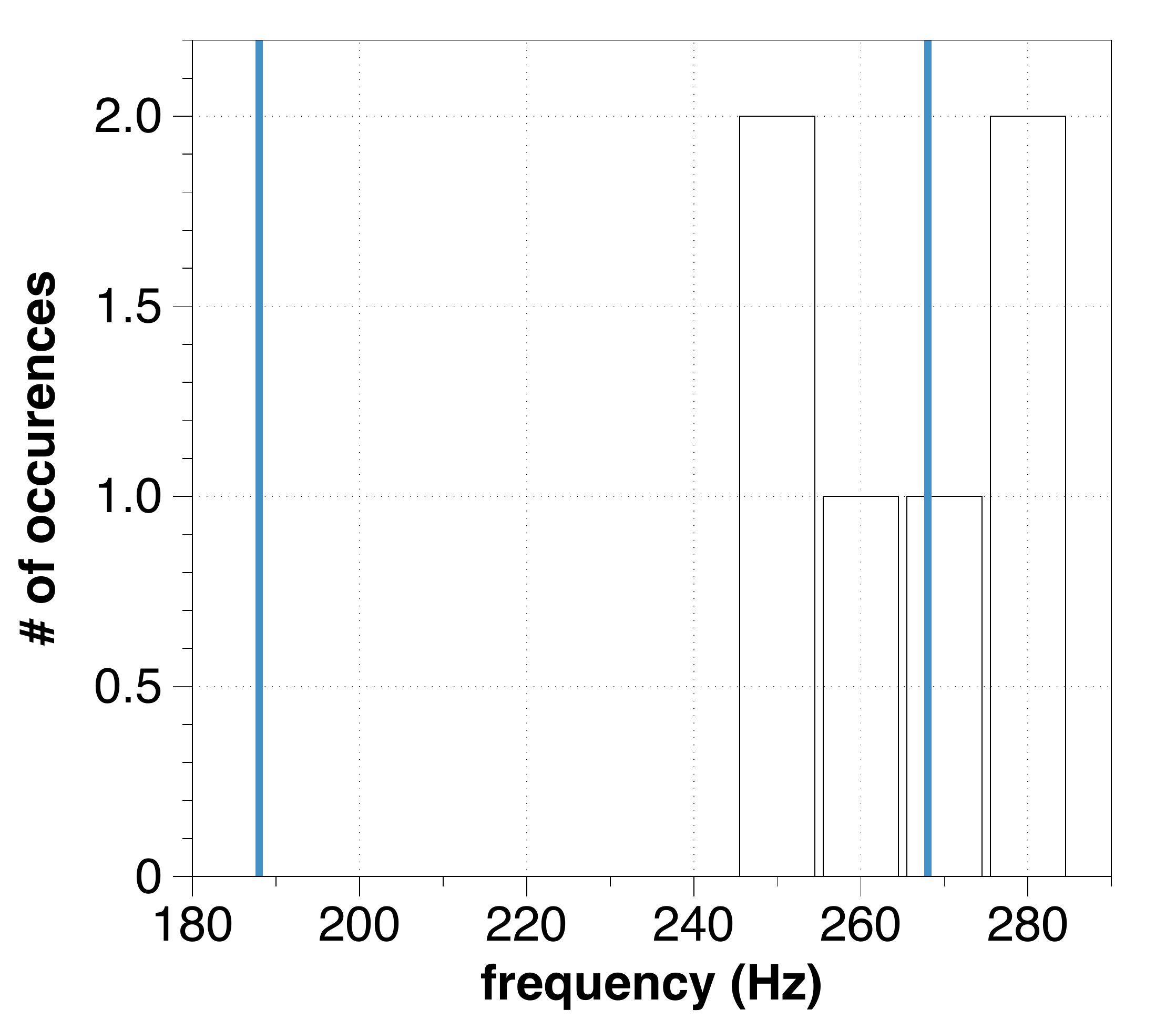}} &
\resizebox{4.cm}{!}{\includegraphics{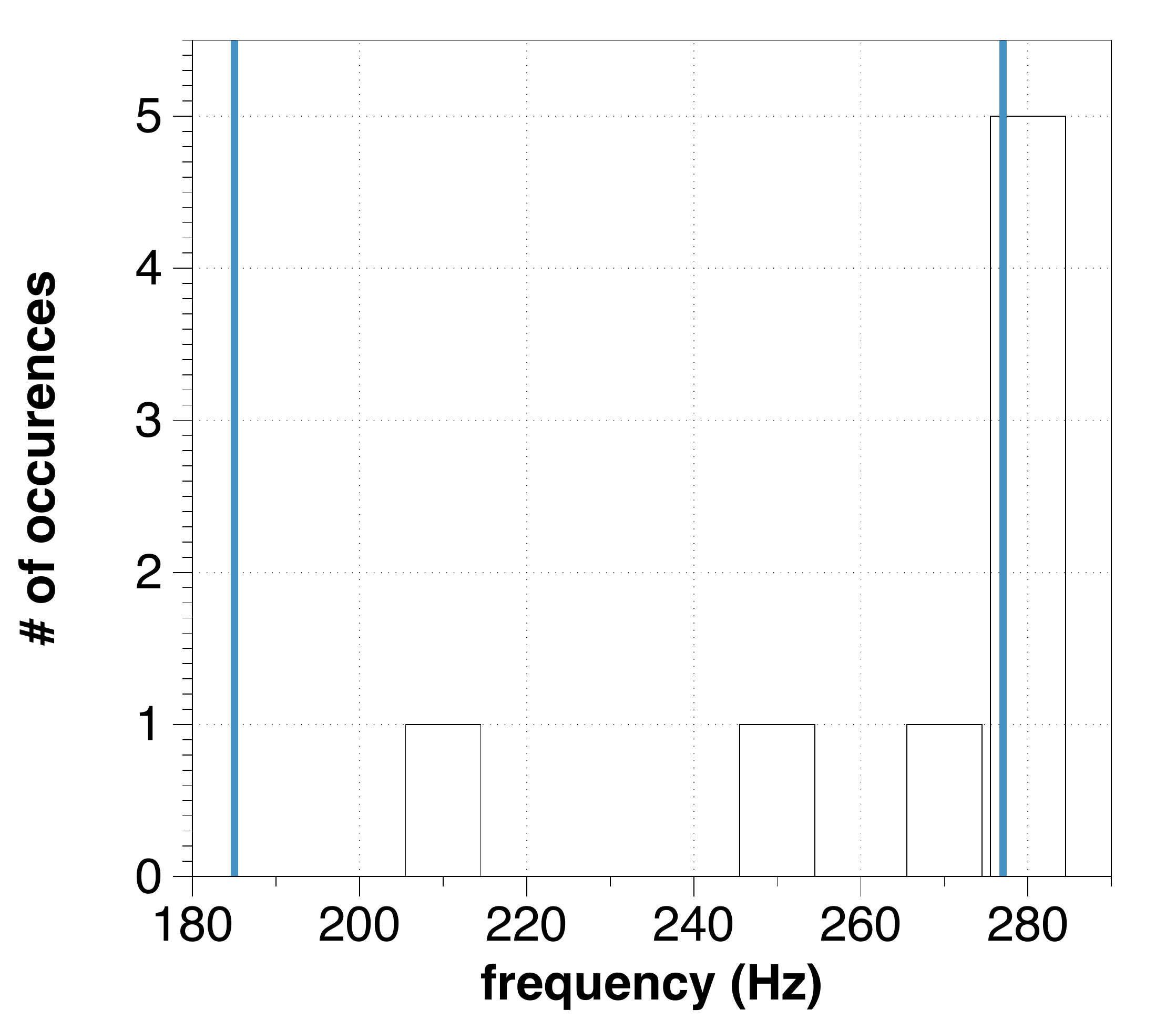}}\\
\end{tabular}
\caption{{\bf{Upper panel}}:  distribution of the HFQPOs' frequencies in \XTE\ during the two outbursts of 1998-99 and 
2000 as function of the ratio of the disk flux to the total flux. The horizontal grey bands represent the groups 
of HFQPOs we  explore in this paper.
{\bf{Lower panels:}} Distribution of the HFQPOs' frequencies in individual observations that were added together by \citet{miller01_1550} on the left 
and \citet{remillard02b} 
on the right. 
The vertical line shows the frequency detected in the averaged PSD.}  
\label{fig:3:2}
\end{figure}

	{In  the upper panel of Fig.\ref{fig:3:2} we report the distribution of frequencies for all HFQPOs detected in 
	\XTE\  \citep{remillard02b,miller01_1550} as function of the ratio of the disk flux to the total flux
	as obtained from spectral fittings \citep{Sobczak00,rodrigue03_1550}. 
 	When looking purely on the frequency level, the distribution} exhibits two  broad \lq clusters\rq, that may be in the 3:2 ratio, although 
	the large variations of frequencies between the different observations, or even within a given cluster, prevent any
	firm conclusion to be obtained.	
	\citet{miller01_1550} produced a global PDS from all observations from the 2000 outburst showing HFQPOs. 
	This total PDS exhibited a pair of HFQPOs, with a main feature  at $268\pm3$ Hz and a a fainter one at $188\pm3$ Hz \citep[][Fig.~\ref{fig:3:2} lower panel]{miller01_1550}. 
	\citet{remillard02b}  separated the observations of the 1998/1999 outburst according to the type of LFQPO they detected. They 
	note an apparent anti correlation between the amplitude of the HF and LFQPO. They also report that broad $\sim$180 Hz QPOs 
	are usually associated with type B LFQPOs, while narrow $\sim$280~Hz QPO are associated with broad type A LFQPO. 
	None of the observations they analyzed showed the presence of a pair of HFQPOs  when analyzed alone. However, accumulating 
	all the (HFQPO) observations with type B LFQPOs on the one hand and those with type A LFQPOs on the other hand, 
	\citet{remillard02b} found that each of the two resultant PDSs exhibited a pair of  HFQPOs, one close to $276$~Hz (respectively 
	$281$ and $277$~Hz) and the other close to $184$~Hz (respectively $187$ and $185$~Hz). Here again the proximity of the 
	frequency ratios with 2:3 led to the conjecture that the HFQPOs are naturally occurring in a $2$:$3$ ratio, a statement further
	 strengthened by the detection, in one case, of the possible fundamental (or the ``1'' of a 1:2:3 ratio at 92~Hz \citep{remillard02b}. \\
	
        As one can see in the lower panel of Fig.\ref{fig:3:2},  the discovery of this ratio was based on adding together many observations 
        with a HFQPO showing a rather wide range of frequencies.  This renders the genuineness and interpretation of this $3$:$2$ ratio 
        difficult as some frequencies are present in individual observations but not in  the averaged data,  while some other reach detectability. 
         One way to overcome the non-detectability by adding more statistics without adding more unknowns is to focus on 
         observations where 1) the already detected frequencies are very close  and 2) the continuum of the PDS, {and the LFQPO complex}
        are also similar. 
        Having a similar continuum means that the behavior/parameters of the inner accretion flow is basically the same 
        and will not influence the PDS by inducing the appearance of extra/{unrelated} components.  
       {This is especially relevant as we showed that  a microquasar's states could be defined by the presence/absence of LF and/or HF QPOs and that these states 
        could be interpreted within a sound theoretical framework \citep{VTR11}.}
        {Said with different words we avoid to include} other sources of variabilities than the one common to all the observations 
        	considered together that could impact the PDS in an unknown manner.
        For the same reason, restraining ourselves to add observations harboring similar HFQPO frequencies will allow us to ensure that 
        the system's parameters are very similar, and that  any other frequencies 
        detected in the averaged PDS is genuinely related to an improved statistics.
        
        When looking at Fig. \ref{fig:3:2} (upper panel), we see that there seems to be several groups of HFQPOs with close 
        frequencies (i.e. within a few Hz from one another; this permits us to define five groups, see Sec.~\ref{sec:reduc},  
        that are represented by the grey horizontal shading in Fig. \ref{fig:3:2}). Using this we will try to 
        answer the following questions: 1) do we detect a {fainter} HFQPO when {combining several observations 
showing similar PDSs and} HFQPOs and in this case 2) {how those fainter HFQPOs relate to the already detected one.}
        With those answers we will be able to probe an eventual relation between the peaks, and in particular check the 
        robustness of the  $3$:$2$(:$1$) ratio.
       
\subsection{RXTE observations and data reduction}
\label{sec:reduc}
We re-analysed all RXTE observations where a HFQPO was reported \citep{miller01_1550, remillard02,belloni12}, and further 
selected those showing multiple (at least two) occurrences of the supposedly same HFQPO. Those are shown as 
squares (hollow red in the online version) in Fig.\ref{fig:3:2}. We considered 
HFQPOs reported from different observations as being the same when 
the difference of their centroid frequencies did not exceed 5 Hz. Of the $\sim$35 reports of HFQPOs 
we then determine five families with distinct frequencies:  $\sim$141~Hz, a $\sim$183~Hz, a $\sim$250~Hz, 
a $\sim$275~Hz, and a $\sim$283~Hz HFQPO, defining what is hereafter referred to as a \lq group\rq\   of data.  
In order to try and increase even more the statistics of our groups, we also considered the {\em{RXTE}} observations found at 
less than 0.5 day from those where HFQPOs were reported if they have a similar continuum, even if they do not show evidence 
for the presence of a high frequency feature.  
Note that \citet{belloni12} also reports the detection of HFQPOs near 263~ÊHz in two observations of the 2000 outburst. 
These features were, however, only detected from the analysis of hard ($>6$~keV) light curves, and since the sample is small 
(amounting a total of 4~ks of data) we decided not include 
them here\footnote{We nevertheless produced a \lq hard\rq\  PDS from these 
two observations, and apart from the already reported 263~Hz QPO \citep{belloni12}, we do not see anything special in this PDS.}. 
\citet{remillard02} also report two occurrences of an HFQPO near 210~Hz. Inspection of the individual PDSs show 
that they are quite different, and therefore we did not consider them as defining a group.
The journal of the observations analyzed in this paper is reported in table \ref{tab:log}.\\
\indent The data from the \rxte /proportional counter array (PCA) were reduced with the {\tt{HEASOFT}} v6.17 
suite following standard procedures to define the good time intervals and 
obtain light curves filtered from data taken at low elevations above the Earth, and large offset from the
source. For each observation, we produced  a high resolution ($2^{-12}$s$=244.140625$ $\mu$s) light curve covering
 the full PCA energy range by combining Single Binned and Event data modes. \correc{These individual light curves were then fed to 
{\tt{POWSPEC}} to produce single-observation power density spectra (PDS, hereafter individual PDS), which were obtained by 
averaging ``sub''-PDSs calculated over temporal intervals of 4~s long. The resultant individual PDSs were, then, rebinned 
 geometrically}. As the source can be 
very bright (up to $\sim13000$ cts/s/PCU) dead time affects the count rates, and in particular artificially lowers the level of Poisson noise 
in the PDSs. The dead time corrected Poisson noise $PN(f)$ was 
estimated for each \correc{individual} PDS by fitting the high frequency ($> 500$ Hz) part of the PDS with a constant. 
Each individual PDS was then corrected from $PN(f)$, and all white noise corrected individual PDSs belonging to a given 
group were averaged following 
\begin{equation}
\mathrm{PDS}_{group}=\frac{\sum_{i} T_i \times {\mathrm{PDS}}_i}{\sum_{i} T_i} 
\label{eq:1}
\end{equation}
\correc{with $group\subset[\mathrm{I, II, III, IV, V}]$,  
PDS$_{i}$ the individual PDS of observation $i$, and $T_i$ the total exposure of observation $i$.}\\

\begin{table}[htbp]
\tablecaption{Journal of the RXTE observations analyzed in this paper separated into the 5 groups of frequencies.}
 \begin{tabular}{ccccc}
 ObsId &  Date  & Rate$^a$ & Exposure & HFQPO  \\
        &  (MJD) &  & (s) & (Hz) \\
\hline
\multicolumn{5}{c}{Group I: $\sim 141$ Hz feature$^\dagger$}\\
\hline
 30191-01-31-00 & 51101.61 & 3107 & 1434  & $141\pm3$ \\
 30191-01-31-01 & 51101.94 & 3058 & 2019 & $145\pm8$\\
\hline
total exposure	& 	& & 3453	& \\
\hline
\multicolumn{5}{c}{Group II: $\sim 183$ Hz feature$^\dagger$}\\
\hline
30191-01-02-00 & 51076.00 & 13223 & 2966  & $183\pm4$ \\
30191-01-33-00 & 51108.08 & 3575 & 9450 & $183\pm3$\\
40401-01-53-00 & 51245.36 & 4607 & 2288 & $182\pm2$\\
40401-01-55-00 & 51247.98 & 4439 & 2462 & $189\pm6$\\
40401-01-56-00 & 51249.40 & 4131 & 542 & $185\pm18$\\
40401-01-56-01 & 51249.47 & 4084 & 180 & $185\pm18$\\
\hline
total exposure	& 	& 	& 17888~s	& \\
\hline
\multicolumn{5}{c}{Group III: $\sim 250$ Hz feature}\\
\hline
40401-01-71-00 & 51270.74 & 733 & 3448 & $253\pm9$$^\dagger$\\
40401-01-72-00 & 51271.41 & 705 & 1976 & \\
50134-02-05-00 & 51669.19 &   958 & 2078  & $249\pm6$$^\ddagger$\\
50134-02-08-00 & 51672.41 &  801 & 1599 & $251\pm6$$^\ddagger$\\
50134-02-08-01 & 51672.96 &  1025  & 1536 & \\ 
50134-01-01-00 &  51673.40 & 962  &  1594 &  \\ 
\hline
total exposure	& 	& 	& 12231~s	& \\
\hline
\end{tabular}
\begin{list}{}{}
\item [$^a$]in units of counts/s per PCU on
\item [$^\dagger$]Report and frequencies of HFQPO from \citet{remillard02}.
\item [$^\ddagger$]Report and frequencies of HFQPO from \citet{miller01_1550}.
\item [$^\star$]Report and frequencies of HFQPO from \citet{belloni12}, from a hard ($\gtrsim6$~keV) band only.
\end{list}
\label{tab:log}
\end{table}

\begin{table}[htbp]
\tablecaption{Journal of the RXTE observations analyzed in this paper separated into the 5 groups of frequencies.}
 \begin{tabular}{ccccc}
 ObsId &  Date  & Rate$^a$ & Exposure & HFQPO  \\
        &  (MJD) &  & (s) & (Hz) \\
\hline
\multicolumn{5}{c}{Group IV: $\sim 275$ Hz feature}\\
\hline
30191-01-36-00 & 51115.28 &   2830 & 2070& $270\pm6$$^\dagger$\\
40401-01-61-00 & 51258.09 &  1927 &  800 & $275\pm 4$$^\dagger$ \\
40401-01-61-01 & 51258.50 &  1882 & 1116  & {276$_{-6}^{+7}$$^\star$} \\
40401-01-62-00 & 51258.97 &  1812 & 1711 &  \\
{40401-01-62-01} & {51259.25} &  {1772} & {735} & {275$\pm4$$^\star$}\\
50134-02-02-01 & 51664.41 & 1500 & 2253 & $276\pm4$$^\ddagger$ \\
50134-02-03-00 & 51664.64 & 1581 & 2870 & $275\pm4$$^\ddagger$ \\
\hline
total exposure	& 	& 	&11555~s	& \\
\hline
\multicolumn{5}{c}{Group V: $\sim 283$ Hz feature}\\
\hline
30191-01-41-00 & 51126.59 & 678 & 4120 & $284\pm4$$^\dagger$ \\
40401-01-50-00 & 51241.80 & 4199 & 3120  & $283\pm2$$^\dagger$\\
40401-01-51-00 & 51242.51 & 4049 & 1517 & $283\pm3$$^\dagger$ \\
40401-01-59-01 & 51255.09 & 2402 & 983 & $281\pm4$$^\dagger$ \\
40401-01-59-00 & 51255.16 & 2369 & 3722 & {281$\pm2$$^\star$}  \\
\hline
total exposure	& 		& & 13462~s	& \\
\hline
\end{tabular}
\begin{list}{}{}
\item [$^a$]in units of counts/s per PCU on
\item [$^\dagger$]Report and frequencies of HFQPO from \citet{remillard02}.
\item [$^\ddagger$]Report and frequencies of HFQPO from \citet{miller01_1550}.
\item [$^\star$]Report and frequencies of HFQPO from \citet{belloni12}, from a hard ($\gtrsim6$~keV) band only.
\end{list}
\label{tab:log}
\end{table}

In all cases the global PDSs have been fitted with {\tt{XSPEC}} V12.9.0i. We  first explored the 0.25--1000~Hz PDS, and fitted 
the full band PDS.  In all cases two or three  zero-centered broad Lorentzians  were used to model the PDSs' continuum. 
Narrower Lorentzians 
(i.e. where Q=$\nu_{\mathrm{QPO}}/\mathrm{FWHM}_{\mathrm{QPO}} \gtrsim2$) 
were then included to represent the low and high frequency QPOs until a satisfactory fit was obtained. 
All errors are given at the 90$\%$ confidence level. To estimate the properties of the HFQPO, we, then, omitted the frequencies below 
50 Hz \citep[see also the discussion in][]{belloni12}. The fit to the continuum is much cleaner and does not depend on a proper 
modelling of the low frequency structures. Simpler models for the continuum were used in this second approach. We verified that 
the parameters obtained for the HFQPOs were compatible with both approaches. As this was the case for all groups, and since our 
focus here is on the HFQPO, we present only the results obtained with the simplest/cleanest approach.\\

\indent \refee{The choice to white noise correct the observations individually first was dictated by the fact that they can have 
very different intensities, and, therefore, 
can be affected differently by dead time effects (in other words PN(f) is observation-dependent). We, however, remark that, in general, for studies of 
HFQPOs, the level of white noise is not subtracted from the PDS, but fitted together with the other components \citep[e.g.][]{miller01_1550,belloni12}. In addition, 
in \citet{belloni12} PN(f) was not consider as constant and was fitted with a power law model with free parameters. To check the consistency of our method
and robustness of the results obtained, we also followed a procedure where all individual, non-PN corrected PDS were averaged following Eq.~\ref{eq:1}, 
and fitted all components in the group PDS thus obtained. PN(f) was fitted with a power law with free parameters \citep{belloni12}. We followed two approaches:}
\begin{enumerate}
\item\refee{We fitted each group PDS over the full frequency range (therefore including the low frequency features), and then kept the best obtained model to 
estimate the parameters of the HFQPO(s), focusing on the frequency range above 50~Hz.}
\item \refee{We only considered the frequency range above 50~Hz (ignoring thus the low frequency complex), to evaluate the continuum (including PN(f)) and 
the parameters of the HFQPO(s).}
\end{enumerate}
\refee{In all cases, PN(f) has a very low power law index (of the order of $10^{-3}$ and sometimes compatible with zero). The parameters of the HFQPOs 
obtained with these two methods are very close (if not identical), and nevertheless always compatible within the (90$\%$) errors, to the one obtained 
with our method, showing the compatibilities of the different approaches. The rest of the paper, therefore, presents and discusses the results obtained with
our method.}

\subsection{Group I: HFQPO at $\sim$141~Hz}
\begin{figure}[htbp]
\centering
\resizebox{9cm}{!}{\includegraphics{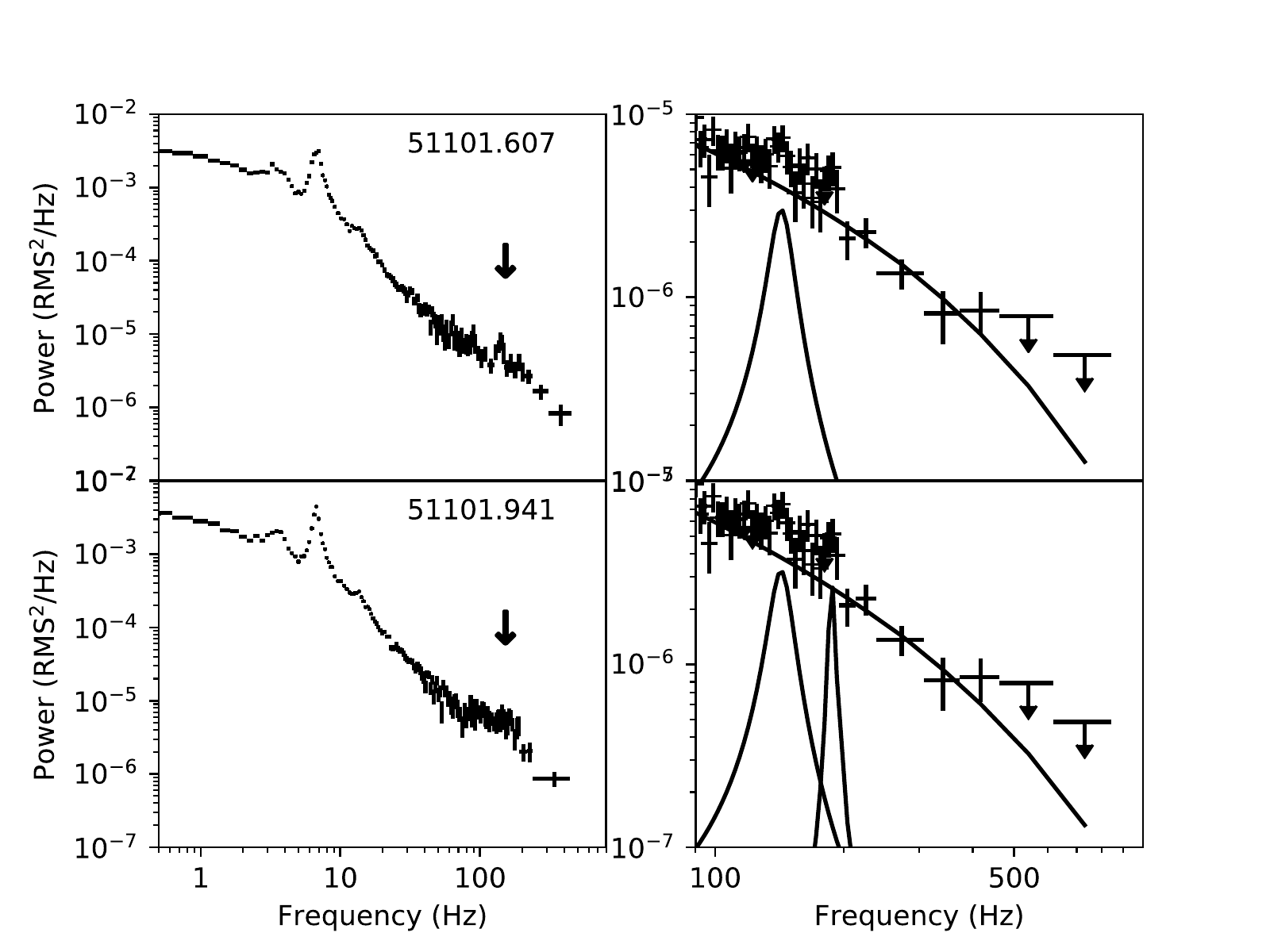}}
\caption{{\bf{Left:}} Individual PDSs of the two observations of group I with a potential 140 Hz HFQPO. 
The vertical arrow shows the approximate location of the HFQPO, which is rather clear in the upper panel.{\bf{Right:}} 
Zoom on the combined PDS from the two 
observations.  The continuous lines represents the individual components of the fit, here a cut-off powerlaw for the continuum, and either one (upper panel) 
or two (lower panel) Lorentzian lines at 141~Hz and 187 Hz.}
\label{fig:140Hz_all}
\end{figure}

Figure~\ref{fig:140Hz_all} (left) shows the individual PDSs of the two observations considered for the analysis 
of the $\sim140$ Hz feature. The HF feature is more  obvious in the first observation, and was reported both in 
\citet{remillard02} and \citet{belloni12}, while the second is only mentioned in the former publication. \\
When omitting the frequencies below 50 Hz, a {zero-centered Lorentzian represents the overall continuum of the combined 
PDS pretty well (\chisq=1.4, 61 degrees of freedom, dof)}. {Some residuals are visible in the the region around 
140~Hz, close to frequency of the previously claimed HFQPOs \citep[][Table~\ref{tab:log}]{remillard02,belloni12}.} 
 The inclusion of a Lorentzian  improves the fit {with $\Delta\chi^2$=20.3 for three additional free parameters. }The width of the feature 
 is not well constrained, and may tend to very broad values. It was frozen to the value reported by \citet{belloni12}, 
{still providing a good fit (\chisq=1.1 for 59 dof). This frozen-width HFQPO is significant at a level of 4.2$\sigma$.} Interestingly some 
excess is still visible around 180--200 Hz (Fig.~\ref{fig:140Hz_all}). Adding another Lorentzian, {however, 
provides a marginal improvement to the fit ($\Delta\chi^2$$\sim$10 for three additional parameters). The additional component is, indeed, statistically 
non-significant (1.7$\sigma$), and we estimate a 3$\sigma$ upper limit of about 1$\%$ for a Q$\sim10$ HFQPO at 
this frequency. We do not consider it further in this study, and take the upper limit estimated here as a proxy for the limit of an additional 
HFQPO for this group.}\\


\subsection{Group II: HFQPO at $\sim$183~Hz}

\begin{figure*}[htbp]
\begin{tabular}{rl}
\resizebox{8cm}{!}{\includegraphics{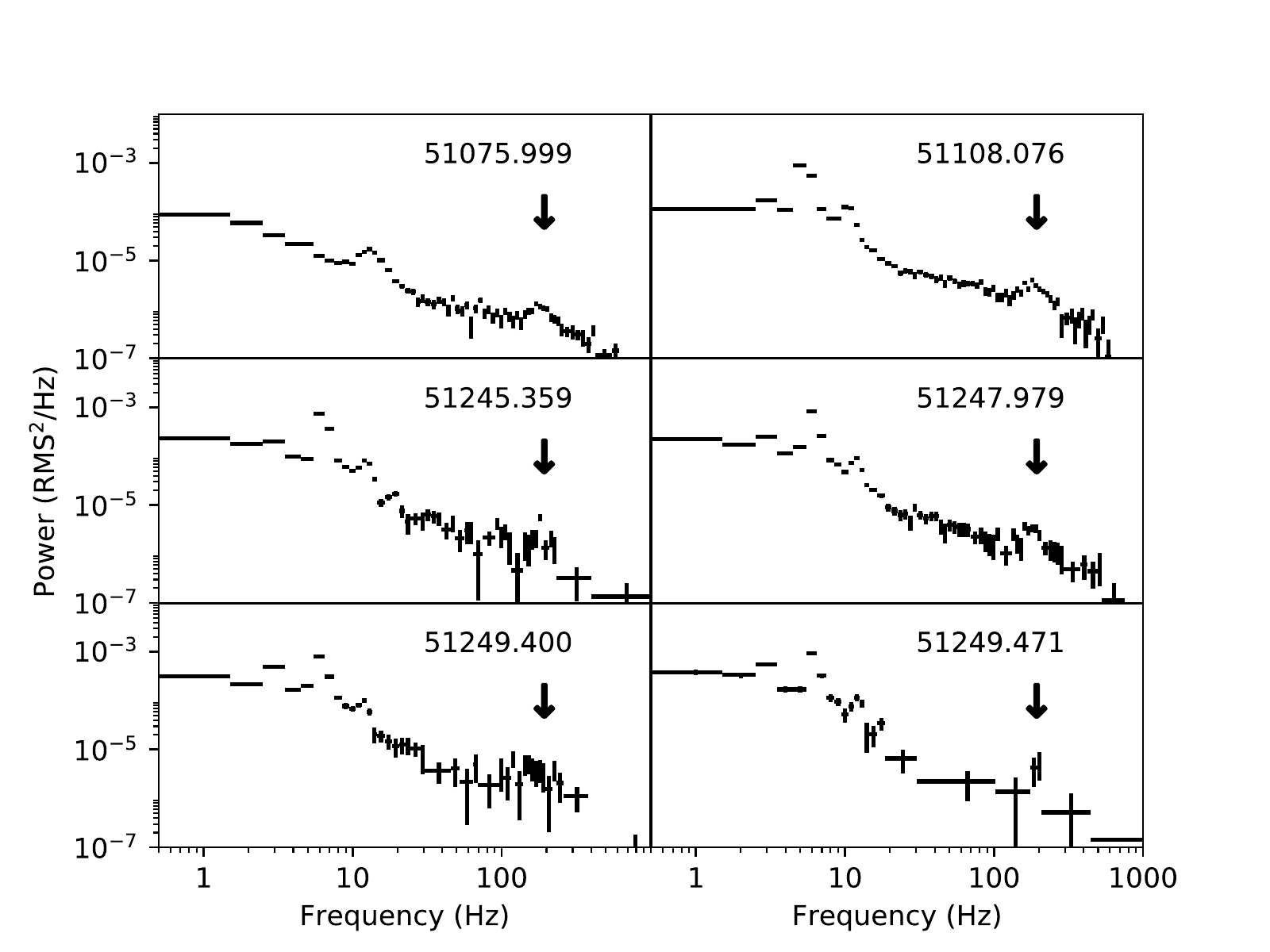}} \resizebox{7.8cm}{!}{\includegraphics{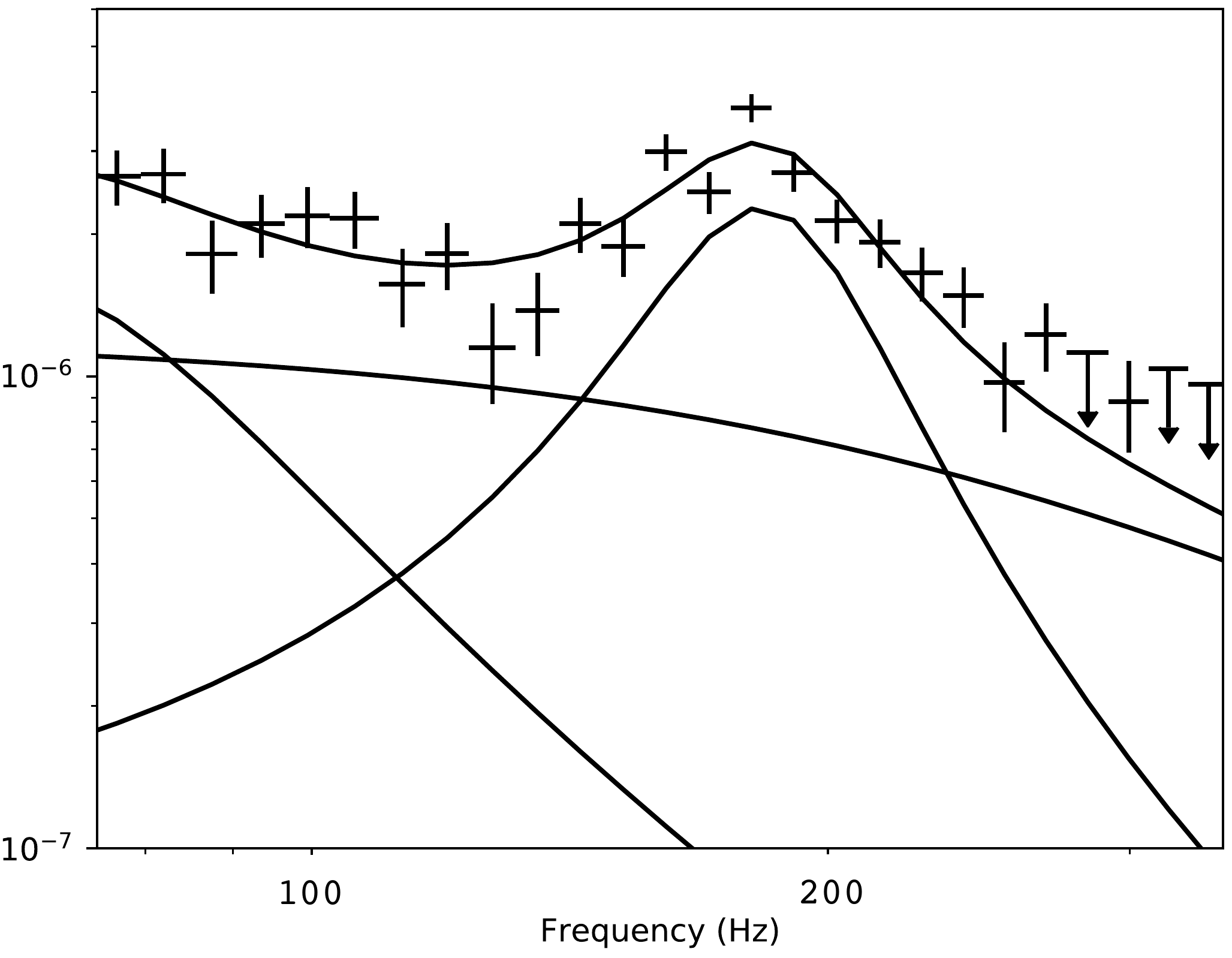}}
\end{tabular} 
\caption{{\bf{Left:}} Individual rebinned PDSs of the six observations  of group II with a potential 183 Hz HFQPO. 
{The vertical arrow shows the approximate location of the HFQPO, which, in all cases, is consistent with an excess}. {\bf{Right:}} {Zoom on the combined} PDS from the six observations. The continuous lines represent the {best fit to the 50--1000 Hz range, and the } 
individual components of the fit. }
\label{fig:180Hz_all}
\end{figure*}

Figure~\ref{fig:180Hz_all} (left) shows the PDSs of all observations considered for the analysis 
of the $\sim183$ Hz feature. We see significant differences in the shapes of the (first and five following) PDSs and in the presence and 
values of the centroid frequencies of the main LFQPO \citep[][indicate variations between 5.4 and 6.4~Hz for the 5 similar 
PDSs, and the LFQPO frequency can go up to 13.2~Hz if we consider the first observation]{remillard02}. 
The LFQPOs are classified as type B by \citet{remillard02}, although those of MJD 51108--51249 show a 
continuum made of ``flat-top'' noise at low frequencies, and rather strong LFQPO complexes with 
(sub)-harmonics that are usually reminiscent of type Cs. 
Despite the different shapes and LFQPO frequencies at low frequencies,  we averaged all PDSs of group II together 
to produce the PDS shown 
in Fig. \ref{fig:180Hz_all}  (right), especially to take advantage of the large photon flux of the MJD 51076.0 observation\footnote{We also considered an averaged PDS obtained while omitting the 
observation of MJD 51076.0, but we obtained no significant differences for the parameters of the combined PDS.}. 
The HFQPO is  visible in at least three of the individual observations and is very obvious in the  the cumulative PDS (Fig.\ref{fig:180Hz_all}). 
The 50--1000 Hz range of the cumulative PDS is best fitted by a model comprised of two broad Lorentzians 
(one zero centered and one centered at about 60 Hz), and a thinner feature, a HFQPO, at 183 Hz. {Although the fit is not perfect (\chisq=1.56 for 
33 dof) attempts to include additional thin features at any frequency failed with estimated upper limits of about 0.8$\%$ for a putative Q=10 HFQPO.}
Of all HFQPOs detected in this study it is the one with the lowest quality factor (Q$\sim$2.9).
We also tested the hypothesis that this broad feature is due to a blend of two narrower peaks, but did not manage to 
obtain satisfactory fits when replacing the broad Lorentzian by two narrow ones even when fixing some of their parameters.

\subsection{Group III: HFQPO at $\sim$250~Hz}

\begin{figure*}[htbp]
\begin{tabular}{rl}
\resizebox{8.cm}{!}{\includegraphics{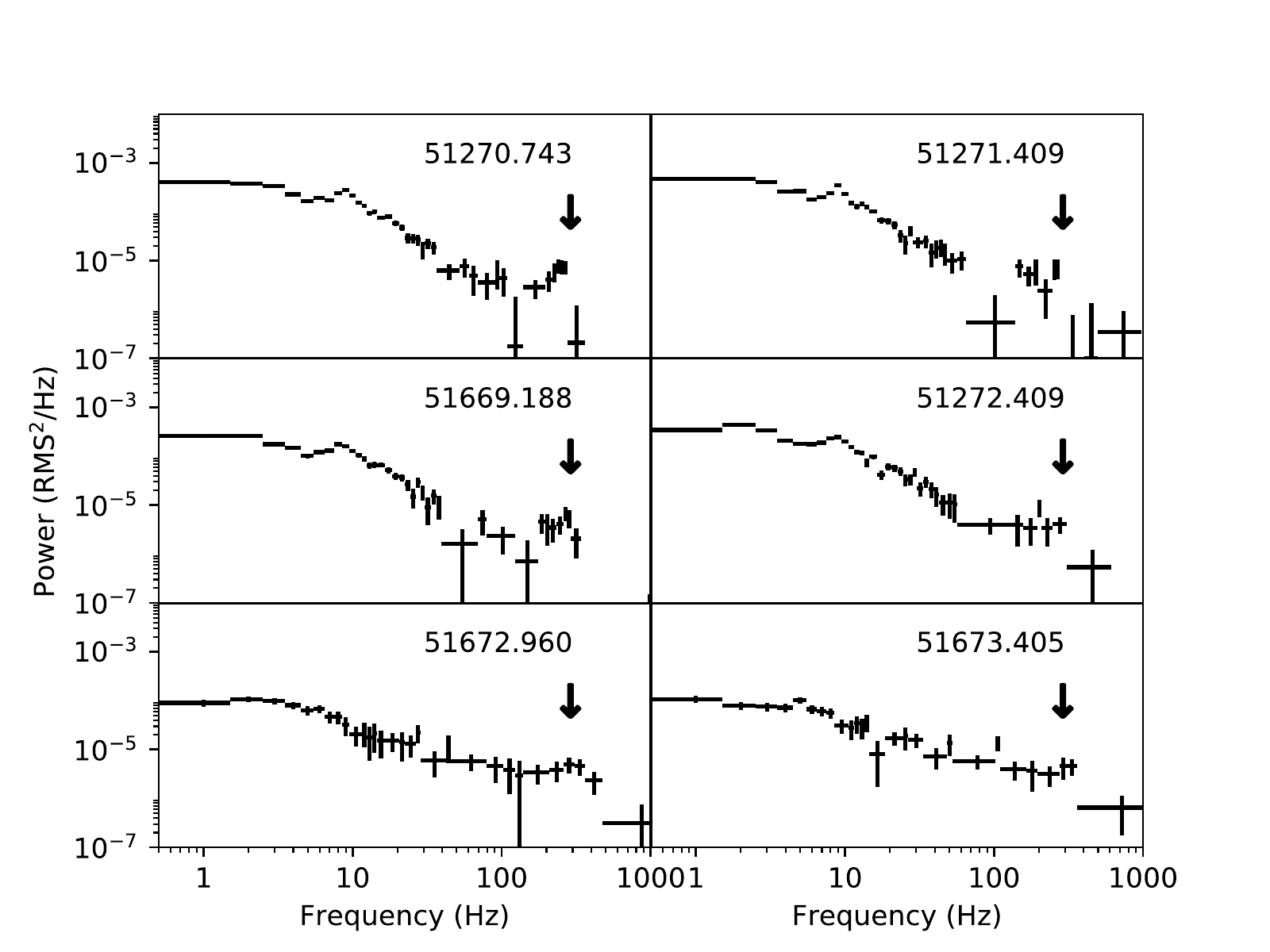}} & \resizebox{7.5cm}{!}{\includegraphics{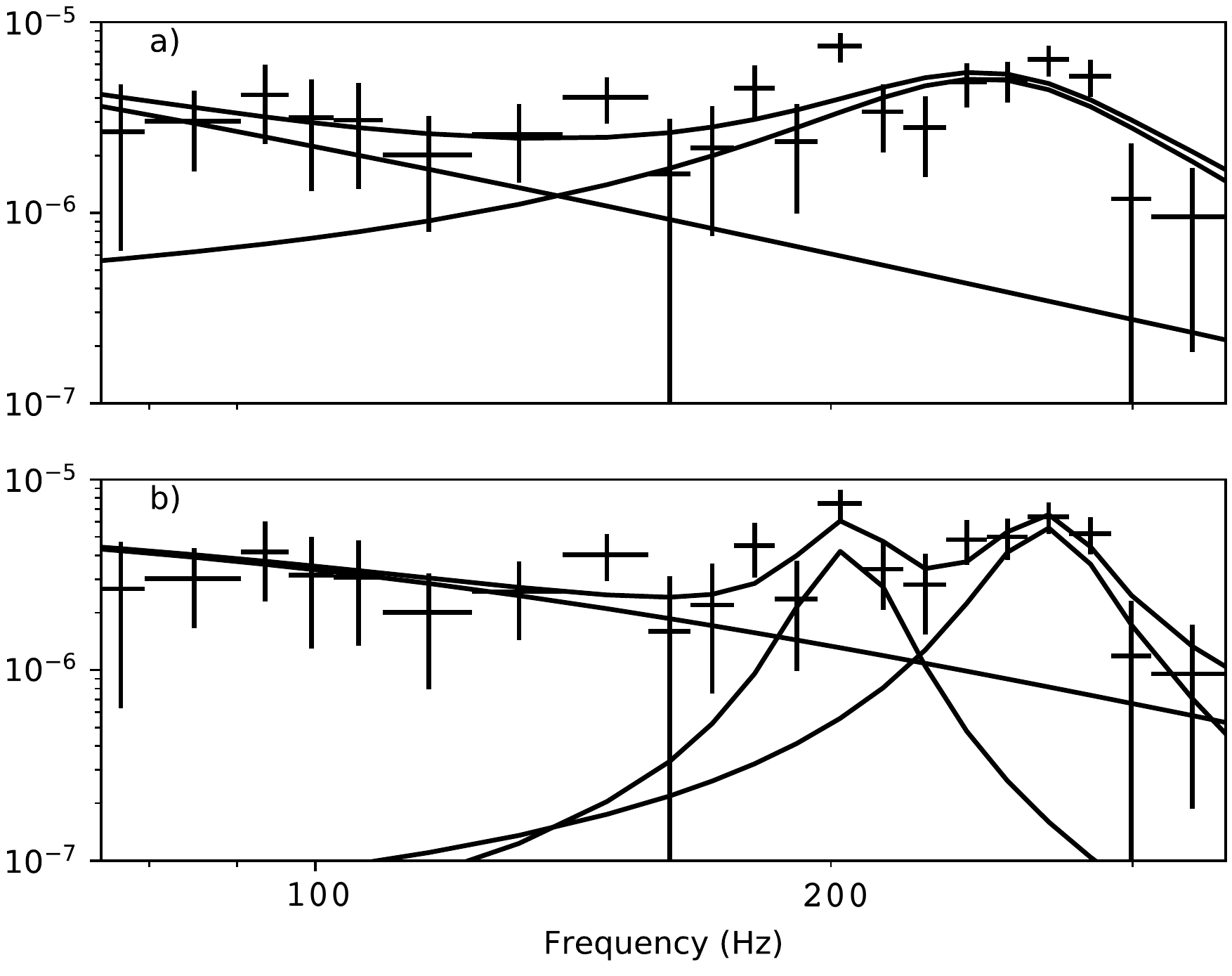}}\\
\end{tabular}
\caption{{\bf{Left:}} Individual {rebinned} PDSs of the six observations of group III considered for the analysis of the 250 Hz QPO. 
{The vertical arrow shows the approximate location of the HFQPO, which is, in most cases, consistent with a slight excess}.{\bf{Right:}} {Zoom on the combined} 
PDS from the seven observations with the best fitted model. a) With a single broad QPO at $\sim245$~Hz. b) With a pair of HFQPOs. 
 The continuous lines represent {the best fit model and} the individual components of the fit. }
\label{fig:250Hz_all}
\end{figure*}

Figure~\ref{fig:250Hz_all} shows the PDSs of all the observations considered for the analysis 
of the $\sim$250 Hz feature. The 250~Hz QPO was reported both during the 1998--99 and 2000 outbursts 
\citep{remillard02, miller01_1550}. It is visible in most of the observations considered here. All individual PDSs are rather similar 
with a low frequency flat top noise and a LFQPO at  $\sim$9 Hz \citep[see also][]{sobczak00b,rodrigue04_1550}. 
The LFQPO is reported as a potential type A with a certain caution in the classification \citep{remillard02} {during 
the 1998--99 outburst, and mentioned as type B during the 2000 one \citep{rodrigue04_1550}. }

When omitting the frequencies below 50 Hz, {the continuum of the combined PDS is  rather well represented with a 
broad zero-centered Lorentzian, although with large residuals in the frequency range of the previously reported QPOs \citep{remillard02,miller01_1550,belloni12}.}
A {thin} additional Lorentzian ($\nu_{\mathrm{QPO}}=245_{-15}^{+22}$~Hz), 
provides a good fit {(\chisq=1.2, 26 dof),  } to the data but the width of the feature tends to very high values that are not compatible with 
a QPO. {A good fit is obtained ({\chisq=1.16, 22 dof}) when fixing the width of the feature to 50 Hz (close to the widths reported in \citet{miller01_1550}).
In this case the best value found for the  QPO frequency is about $260$ Hz (Table~\ref{tab:qpofit}). }\\
\indent {In an attempt to verify if the tendency of the fit to converge toward a broad feature was not due to the presence of 
two HFQPOs, and since in the fixed width fit some residuals are still visible at about 200 Hz, we, in a second run included 
a second thin Lorentzian  to the model}. {The fit quality marginally improved ($\Delta\chi^2$$=$13 for 3 additional dof), 
with a final \chisq$<$1 indicating that the model oversamples the PDS. The width of the second component is unconstrained, and fixing 
it to 50 Hz permits a good fit to be obtained with a \chisq close to one (25 dof). }The statistical significance of the second 
feature is  3.1$\sigma$. {Given this rather low significance, and the marginal improvement 
brought by the inclusion of this second putative feature, one can question its genuineness, and we prefer to consider it as a possible 
second QPO rather than a soundly confirmed one. Bearing this in mind, the}  
ratio of the frequencies of the two HFQPOs  ($f_{\mathrm{low}}/f_{\mathrm{high}}$) {would then be}  $0.75^{+0.02}_{-0.03}$.  
{It is interesting to remark that the centroid frequency of the main QPO (265 Hz) is almost exactly the value of the \lq hard\rq\  263~Hz 
QPO reported in two occasions by \citet{miller01_1550}\footnote{Dates are reported with a one day error in this paper} and \citet{belloni12} from a hard 
band only. }

\indent {In order to check that this  occurrence of a potential pair of HFQPO was not due to a mixing of observations with different QPOs (although none 
other than the 250 Hz were previously reported from the observations considered in this group), we produced a PDS  from the three  observations 
where the feature was previously reported \citep[Table~\ref{tab:log}]{miller01_1550,remillard02}. The resulting PDS (not shown) is well fitted (above 50 Hz) 
by a broad zero-centered Lorentzian plus a thin one (\chisq=0.9 for 23 dof). The width is however unconstrained and the centroid found at 
244($_{-17}^{+14}$) Hz. We then froze the width to 50~Hz \citep{miller01_1550}, and obtained a satisfactory fit (\chisq=0.9, 24 dof) with 
$\nu_{\mathrm{QPO}}=250\pm11$~Hz, and P=$2.2_{-0.4}^{+0.3} \%$ RMS}. {It is significant at 5.1$\sigma$. We estimate a 3$\sigma$ 
limit of  about 1.5\% (resp. 1.8 \%) for the presence of an additional Q=10 (resp. Q=4) feature at 200~Hz}. \\
\indent {Similarly, we then considered a PDS averaged over the three observations where no QPO was previously reported (Table~\ref{tab:log}). 
The PDS is well represented by a single broad zero-centered Lorentzian (\chisq=0.9 for 24 dof). We estimate a 3$\sigma$ limit of  about 1.9--2.0$\%$ 
for a 50~Hz FWHM HFQPO at 200--260~Hz.}

\subsection{Group IV: HFQPO at $\sim$275~Hz}

\begin{figure*}[htbp]
\begin{tabular}{rl}
\resizebox{8.3cm}{!}{\includegraphics{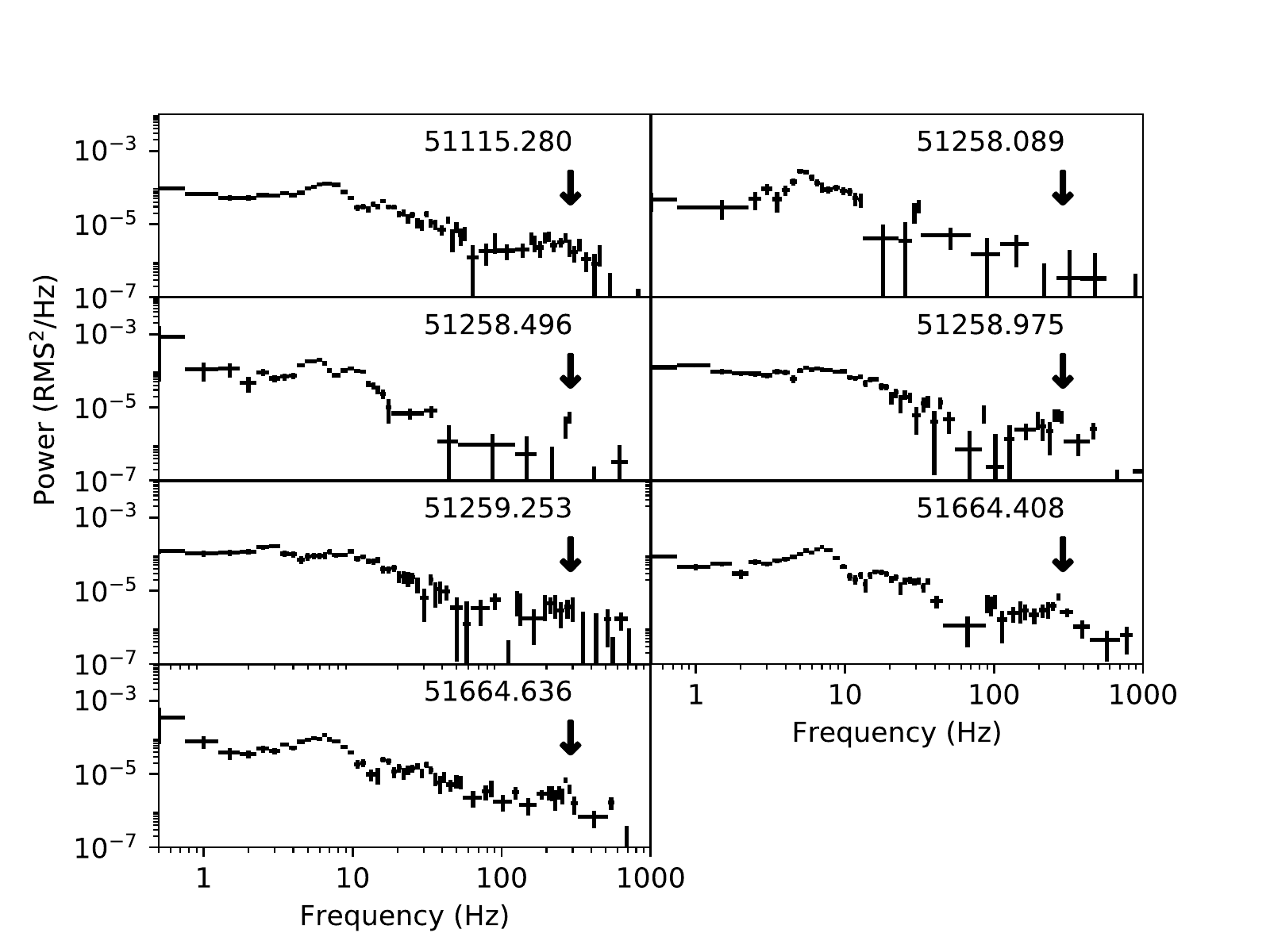}} & \resizebox{8cm}{!}{\includegraphics{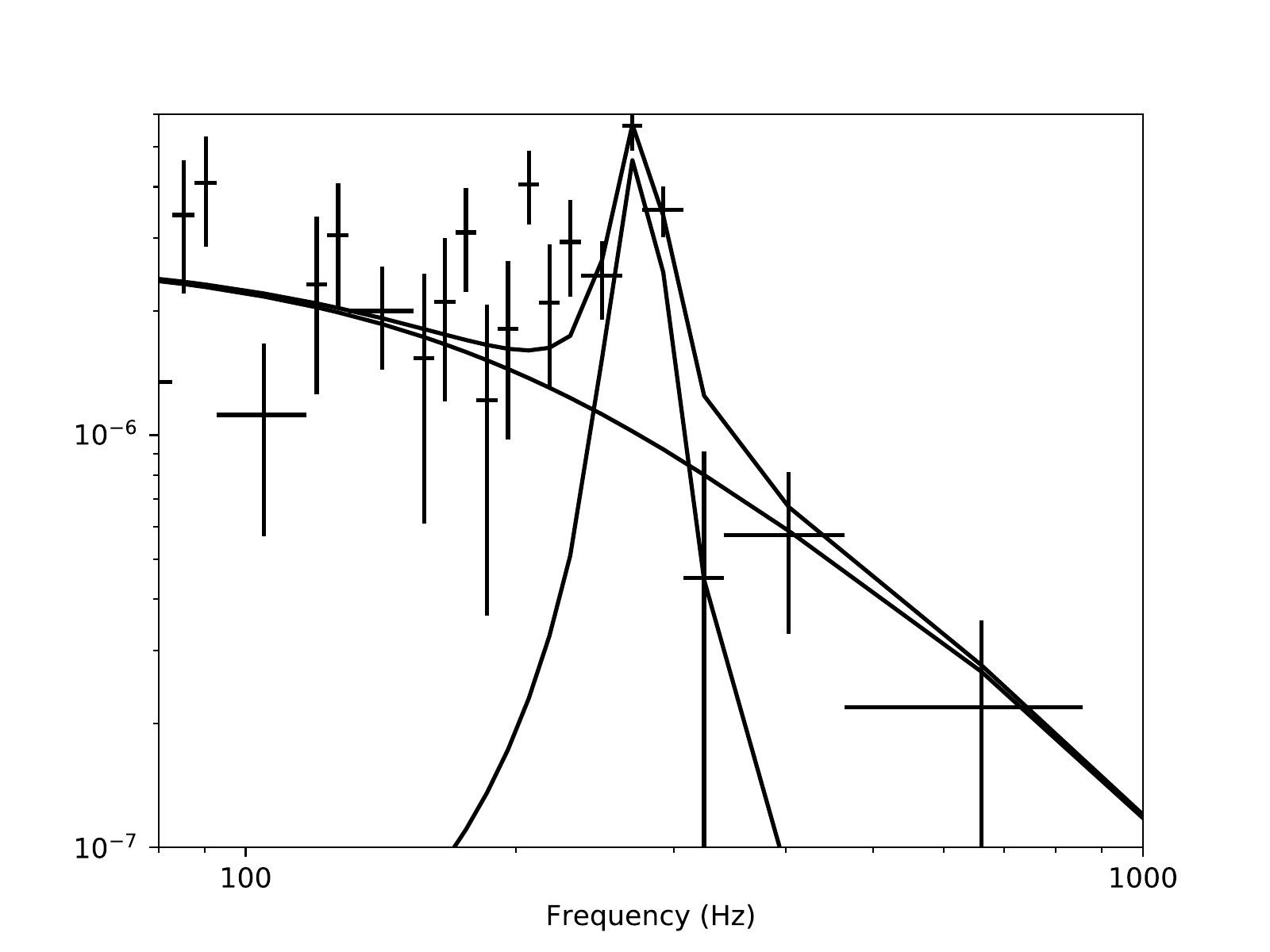}}\\
\end{tabular}
\caption{{\bf{Left:}} Individual {rebinned} PDSs of the six observations of group IV  with a potential 275 Hz HFQPO. 
{The vertical arrow shows the approximate location of the HFQPO, which is, in most cases, consistent with a slight excess}.{\bf{Right:}} {Zoom on the 80--1000 Hz region of the }combined PDS from the six observations.  The continuous lines represent the {best fit model} and the individual components 
of the fit.}
\label{fig:275Hz_all}
\end{figure*}

Figure~\ref{fig:275Hz_all} shows the PDSs of all the observations considered for the analysis 
of the $\sim275$ Hz feature. As for the 250~Hz, the HFQPO near 275~Hz was reported in both the 
1998--99 and 2000 outbursts \citep{remillard02, miller01_1550,belloni12}.
All individual PDSs are roughly similar, with a flat top noise component and a broad low frequency feature  (at about 
5.5~Hz) on top of it with a potential harmonic around 10 Hz \citep[see][]{sobczak00b,rodrigue04_1550}. The LFQPO is reported as  
a type A in the MJD 51115.28 observation and a possible type A with a certain caution in the classification in the MJD 51258 
observations by \citet{remillard02}. \\
\indent When omitting the frequencies below 50 Hz, the combined PDS is rather well represented {(\chisq=1.3 for 17 dof) 
with a broken power law continuum (with a rather flat first component and a break at about 230 Hz) and an additional thin feature 
at $\sim275$~Hz  (Fig.~\ref{fig:275Hz_all} right, Table~\ref{tab:qpofit}). This model is preferred to a broad zero-centered Lorentzian plus 
the QPO which gives a poorer fit (\chisq=1.6 for 19 dof) or just a broad Lorentzian plus the QPO (\chisq=1.5 for 18 dof).} 
{As the model of the underlying continuum influences the results of the QPO we nevertheless estimated the parameters of the 
QPO with the three models. In all cases, the frequencies obtained are the same, with similar errors, the width is slightly lower in the broken powerlaw
case even if all widths are consistent within the large errors, and the RMS amplitude is also slightly higher in the zero-centered case ($1.5\pm0.2\%$) 
than in the broken power law fit ($1.2\pm0.3\%$, Table~\ref{tab:qpofit}), although all values are compatible within the 90\% errors. We thus consider in 
the following the broken powerlaw results only.}
This HFQPO {is significant at about {6$\sigma$}. 
No other peak is obvious in the combined PDS and we estimate 3$\sigma$ upper limits  comprised between $\sim1\%$  and 1.8\% for the presence of 
an additional HFQPO in the considered range.}

\subsection{Group V: HFQPO at $\sim$283~Hz}

\begin{figure*}[htbp]
\begin{tabular}{rl}
\resizebox{8.2cm}{!}{\includegraphics{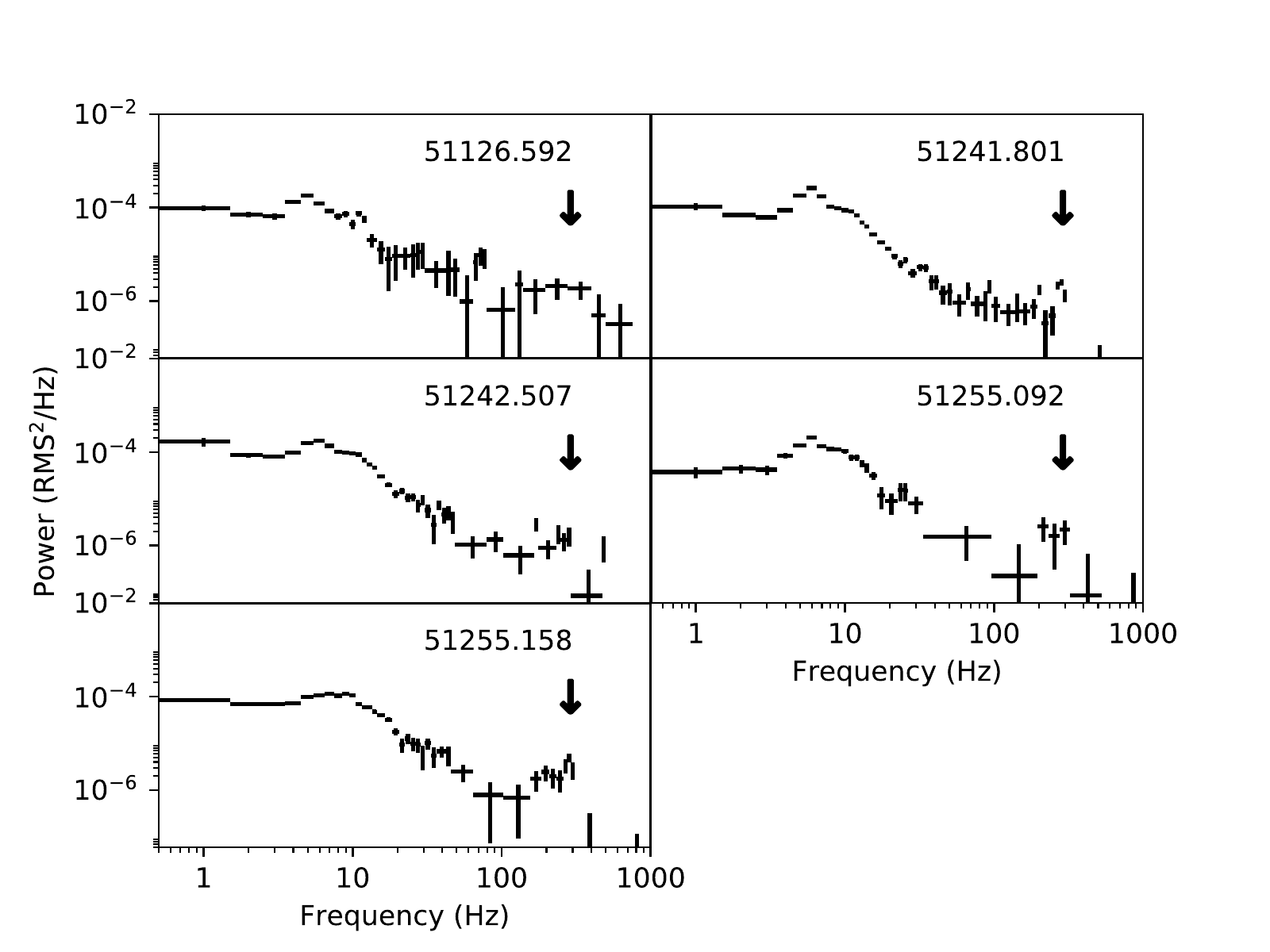}} & \resizebox{8cm}{!}{\includegraphics{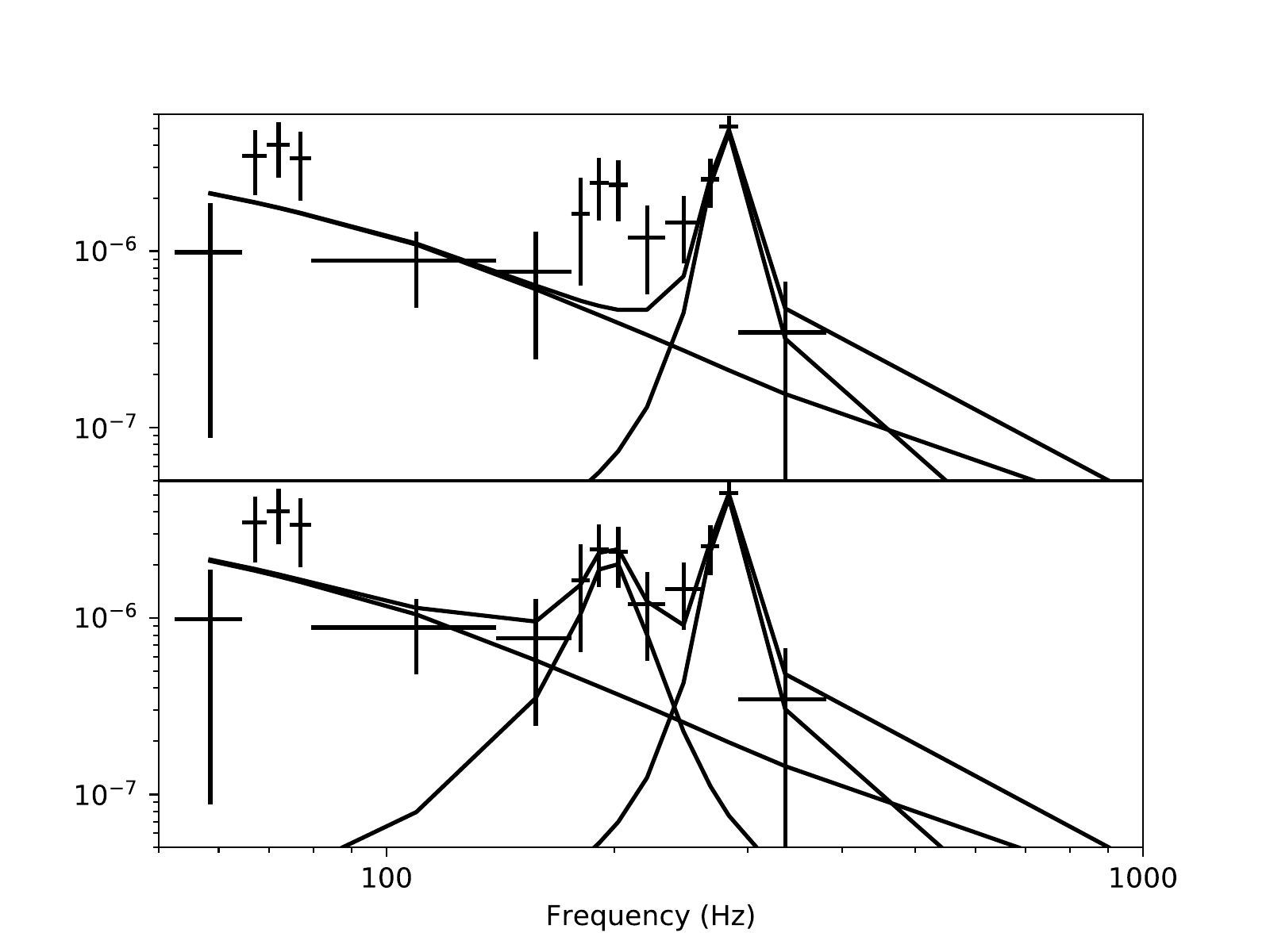}}\\
\end{tabular}
\caption{{\bf{Left:}} Individual {rebinned} PDSs of the five observations of group V considered for the analysis of the 283 Hz QPO. {The vertical arrow shows the approximate location of the HFQPO, which is, in most cases, consistent with a slight excess}.
{\bf{Right:}} {Zoom on the 50--1000~Hz range of the }combined PDS from the five observations with the best fitted model, that includes a pair of HFQPOs.  The continuous lines represent the individual components of the fit.}
\label{fig:283Hz_all}
\end{figure*}

Figure~\ref{fig:283Hz_all} shows the PDSs of all the observations considered for the analysis 
of the $\sim$283 Hz feature. {When omitting the frequencies below 50 Hz,} the continuum of the combined 
PDS is well modeled by one broad zero centered Lorentzian. This combined PDS clearly shows the presence of a peak 
at $\sim$280~Hz, {and adding a thin Lorentzian provides a rather good fit (\chisq=1.5 for 21 dof) to the PDS,  but large residuals still remain}. 
{The inclusion of a second narrow Lorentzian {slightly} improves the fit ({\chisq=1.2 for 18 dof, ie} 
$\Delta\chi^2=10$ for 3 additional degrees of freedom). 
The width of this extra component is, however, not constrained, and tends to very high value especially when the 
continuum is not fixed. This effect is clearly due both to the restrained frequency range of our fit, and the three bins around 
60--80 Hz that tend, here, to be considered as a thin feature (while there are not when considering the broad band PDS). In a second 
pass we froze its width to the centroid of the best value obtained when left free (Table~\ref{tab:qpofit}).} 
In both the fits with or without the second peak, the 283~Hz QPO is significant at {7$\sigma$}. 
We estimate the significance of the second feature to be {just at the 3$\sigma$ level}  (Table~\ref{tab:qpofit}).
{Here again, given the rather low significance of the second potential QPO, and the marginal improvement 
brought by the inclusion of this second putative feature,  we prefer to consider it as a possible 
second QPO rather than a confirmed one. Bearing this in mind, the}  
ratio of the frequencies of the two HFQPOs  ($f_{\mathrm{low}}/f_{\mathrm{high}}$) {would then be $f_{\mathrm{low}}/f_{\mathrm{high}}$$\sim 0.72^{+0.04}_{-0.07}$}\\

\section{Discussion}
\subsection{{Is the lack of multiple peaks due to the limited statistics of some PDSs}?} 

\begin{table}[!t]
\caption{Parameters of the HF QPOs obtained from the fits to the stacked PDSs of each group of data. {The errors correspond to the 
90\% level of uncertainty.}}
\begin{tabular}{lccccc}
Group & HFQPO & FWHM & Power  &  significance$^a$ \\
               &  freq (Hz)              &     (Hz)  &    (\% RMS)   & \\
\hline
I  ($\sim$ 140Hz)  & $143\pm4$  & 16.5$^c$ & $0.9\pm0.2$  	&  $4.2\sigma$ \\
\hline
II ($\sim$ 183Hz)  &     $182\pm4$  &  $64_{-21}^{+24}$ & $1.5\pm0.3$    & $11\sigma$  \\
\hline
III$^d$($\sim$ 250Hz)        &     $260_{-11}^{+10}$  &  $50$$^c$ & $1.9\pm0.3$ & $\sim5\sigma$ \\
\multicolumn{5}{r}{  --  --  --  --   --   --   --   --   --   --   --   --   --   --   --   --   --   --   --   --   --   --   --   --   --   --   --   --}\\
III$^e$ ($\sim$ 250Hz)        &     $267\pm8$  &  $50^c$ & $2.0\pm0.3$  &  $\sim5.9\sigma$ \\
                   &     $200_{-13}^{+11}$ & $50$ frozen & 1.6$_{-0.5}^{+0.3}$ & $3.1\sigma$ \\ 
\hline
IV ($\sim$ 275Hz)            &   $276_{-4}^{+2}$    & 10$_{-8}^{+25}$  &   $1.2\pm0.3$   &  $5.9\sigma$   \\

\hline
V   ($\sim$ 283Hz)          &   $280_{-4}^{+5}$    & 19$_{-13}^{+15}$  & $1.3\pm0.2$   &  $\sim7\sigma$  \\
                &   $202_{-14}^{+9}$    & $20$ frozen  & $0.9_{-0.3}^{+0.2}$  &  $\sim3\sigma$ \\
\end{tabular}
\begin{list}{}{}
\item[$^a$] {Significance calculated for the best fitted, or for a fixed width}.
\item[$^b$] Parameter pegged at the upper limit for the feature to be considered a QPO.  
\item[$^c$] Parameter frozen to the common value found in the literature.  
\item[$^d$] Group III fitted with only one broad HFQPO (Fig.~\ref{fig:250Hz_all}a).  
\item[$^e$] Group III fitted with two HFQPOs (Fig.~\ref{fig:250Hz_all}b).  
\end{list}
\label{tab:qpofit}
\end{table}
 
     {Our first aim here was to check if multiple peaks are present but not detected because they are below the detection level 
     in the single observations.  Our definition of the groups permitted us to significantly extend the total duration of each considered
     sample, and thus lower the level of detectability for thin features in the combined PDSs (Table~\ref{tab:log}) except for group I.}
     {In our five groups we get  two potential ($3\sigma$ level) pair detection in group III and group V (Table \ref{tab:qpofit}), while the very
     short exposure (3.5 ks) of group I does not permit to establish the reality of the 1.7$\sigma$ peak mentioned above.  } 
     {The absence of detection of any additional peak(s) (other than those already reported  in the literature) in groups  II and IV,
     clearly indicated that increased statistics does not always lead to the detection of multiple peaks.}
          {Our best  example is group II which has the longest total exposure, of about 18 ks, but shows only one broad peak with a 3$\sigma$ 
       fractional RMS limits  of 0.8\% for the presence of an additional  HFQPO.}

{This in turn raises another question. What are the conditions that trigger the presence{, or more precisely the detectability,} of more than one peak? 
While fully answering this question
is beyond the scope of this paper we can look at some directions for future studies.}
Interestingly, while group IV and V have rather similar PDSs, one shows a {potential} pair of peaks and the other does not. 
The fact that group IV has a HFQPO similar to that of  group V, but no other peak detected suggests that other 
parameters have an impact on the number of significant HF features  in the PDSs. They  could, for example, be influenced 
by the overall continuum shape and power,  or the full LFQPO structure, that 
shows, here, strong differences {as we briefly mention here in the introduction of each group, and as 
is obvious from the in-depth analysis of the 1998--1999 and 2000 outbursts \citep{remillard02,rodrigue04_1550}}.   \\

\begin{figure}[htbp]
\centering
{\resizebox{8.5cm}{!}{\includegraphics{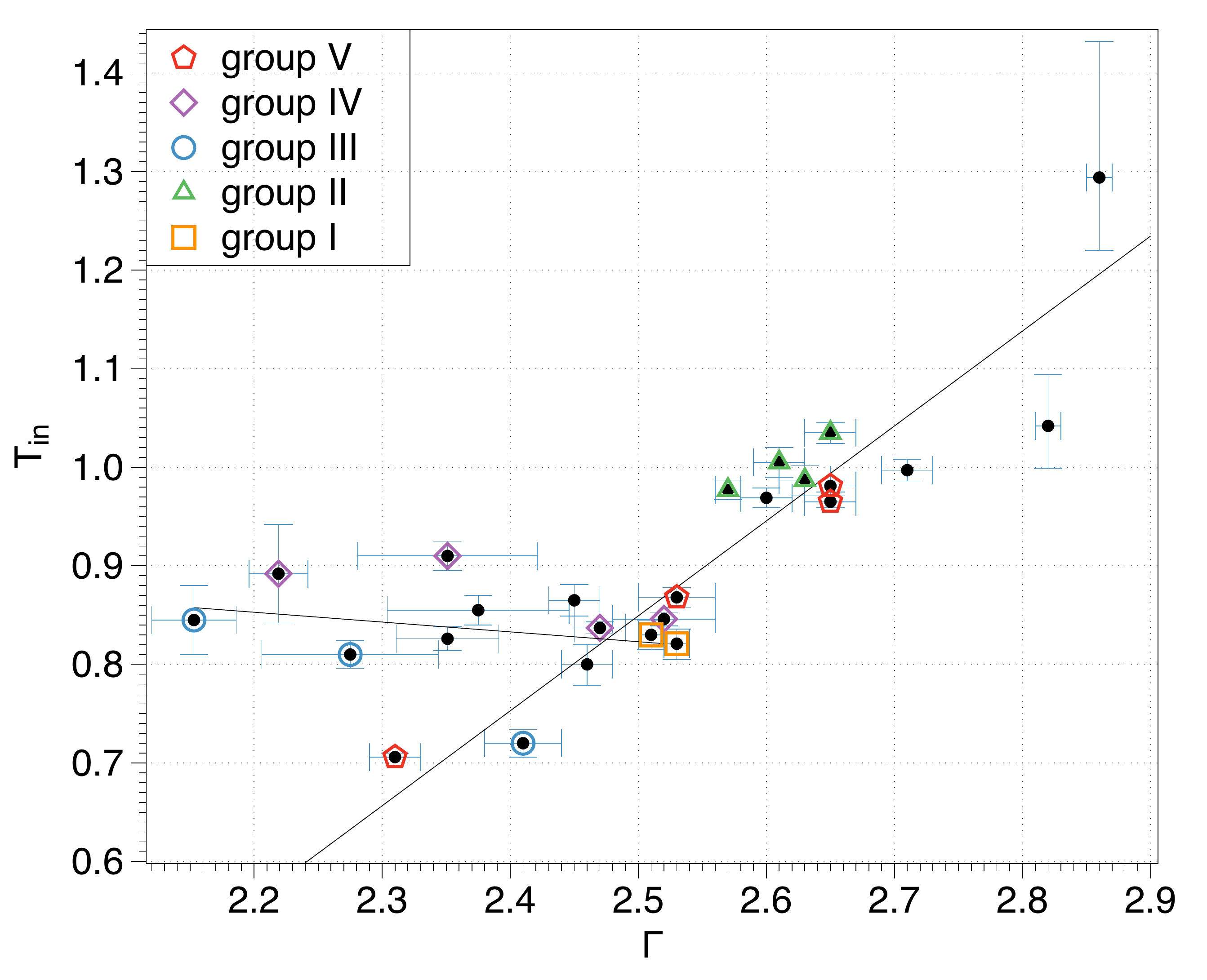}}}\\
\caption{  {Representation of the behavior of the inner temperature $T_{in}$ versus the Photon index $\Gamma$ for all the HFQPOs detected in the
98-99 and 2000 outburst as in \citet{miller01_1550, remillard02b}. Group I is represented by the hollow orange square, group II the hollow green triangle,
group III the hollow blue circle, group IV the hollow purple diamond and group V the hollow red pentagon.
}  }
\label{fig:Gamma_Tin}
\end{figure}	
	            
 \indent Another possibility is a difference of the spectral behavior of the source between the different groups. 
 While a   spectral analysis of the five groups we defined is not available and beyond the scope of this paper, we can look at the 
 source behavior during the individual observations. Direct comparison between the HFQPO frequency and
 the power law  photon index ($\Gamma$)  of the energy spectra \citep{Sobczak00, rodrigue03_1550}
 does not show any particular relation {but groupe IV  has a power law  photon index between   2.22 and 2.52
  $\pm 0.04$ while group V is slightly higher with $\Gamma$ between 2.31 and 2.65 $\pm 0.02$}.  
 {Similarly,  group V exhibits a wider range of inner temperature ($T_{in}$ between 0.71 and 0.98  $\pm 0.006$) 
than group IV (between 0.837 and 0.910 $\pm 0.015$).
When putting those together the behavior of group IV and V in the $(\Gamma,T_{in})$ space is perpendicular to one-another (see Fig \ref{fig:Gamma_Tin}).
Altogether, this shows that both groups have a different spectral behavior which might be the cause of only one group having a second detectable peak.}\\

\subsection{Status of the $3$:$2$ ratio in our dataset}

{Now that we have shown that {stacking together PDSs of similar HFQPO observations, can sometimes lead to a second, tenuous ($3\sigma$),
peak detection,}   we can study how the detected frequencies}  relate to one another.
 In  group V we have a {peak ratio $\nu_{1,\mathrm{HF}}$/$\nu_{2,\mathrm{HF}}$=$0.72^{+0.04}_{-0.07}$}
{which is compatible with several integer ratios, thus closer to being 3:4,  while the 2:3 ratio is at the boundary 
of the errors.}
If we look at group III the ratio of observed frequencies is { $\nu_{1,\mathrm{HF}}$/$\nu_{2,\mathrm{HF}}$=$0.75^{+0.02}_{-0.03}$
which is close to the ratio 3:4 but incompatible with a 2:3.}
So while the {potentially} detected pairs of frequencies could be in some integer relationship with one another, we did not detect a {\bf firm} 2:3 
ratio between the peaks but, we instead obtained {preferably} higher integer ratios in the two cases where we have detected 
a {possible} pair of HFQPOs. 
We {can therefore {simply} 
conclude that {the occurrence of HFQPOs in pairs with specific or preferred frequency} ratio (in particular the 3:2) 
should not be taken as a strong requirement for HFQPO models. }\\

\indent {Let us now now consider the possibility that all the (HFQPO) frequencies 
exhibited by XTE J1550$-$564  relate to one another in an integer ratio (ie that they are somehow all harmonically related). 
  While this assumption may first appear counter-intuitive there are several facts pointing in that direction.
  First of all, the fact that the detected frequency can be relatively similar for observations that are separated by years \citep[e.g.][]{belloni_alt2013},
  points toward a mechanism able to trigger the 
  same frequencies at very different, non connected moments (in particular different outbursts). 
  Second the, now, {potentially} detected 
  simultaneous occurrence of pairs, indicates a mechanism able to stimulate multiple frequencies at the same time. 
  So, assuming that all the XTE J1550$-$564 HFQPOs are indeed a manifestation of the same mechanism we can associate 
  the HFQPO frequency of the different groups to harmonic numbers. In this exercise, one would need to allow for a much 
  lower base/fundamental frequency and much higher ratio and number of harmonics, to fully represent the whole sample.}
  For example, the detection of HFQPOs at 141 Hz, 198 Hz, or 281~Hz, would then require to go up to mode 9 or 10 with a 
  fundamental frequency around 30--35~Hz. \\
  \indent While this cannot be totally excluded, it rather raises extra difficulties and questions regarding, for example, 
  the main selection criteria for these different modes. 
While we do not favor this interpretation, it is interesting to note that such a high number of modes is sometimes seen in fully General Relativistic disk simulations 
\citep[Casse \& Varni\`ere 2018, submitted]{CV17} but rarely in the Pseudo-Newtonian case \citep{Vin13}.  \\

\indent Another way to reconcile those observations with the peaks being harmonically related without the need to have high 
modes is to relax the assumption that  the frequency of the {fundamental} is mostly constant. 
So, depending on the physical condition in the system, we keep a relationship between the peaks up to four with 
a base frequency that {can} vary by a factor of up to two. 
This interpretation, however, has its own set of difficulties, mainly related to the origin of the  limited variation of this base frequency. 
 
 Another possibility is that our limited sample is not representative of the full extent of the variation of the HFQPO frequency, 
    or the set of frequencies are related to a single peak,  in which cases the models need to incorporate an ever larger range of variation. \\

{In conclusion, by increasing the total exposure of five HFQPOs' frequencies, we did not detect another peak for three groups, while 
also obtaining two $3-\sigma$ potential detections of a second peak, neither of which favored a 3:2 ratio. }


 
 \subsection{Constraints for HFQPO models}
 
 \indent Using the known facts and the results obtained here one can form a list of contraints that any HFQPO model needs to explain.
 
 \begin{itemize} 
    \item an HFQPO model needs to modulate the X-ray flux. \\
    This is of course an obvious fact, and might be the most straightforward constraint but it is not often addressed by the theoretical models.\\
     
     \item A single mechanism should be able to predict the {distribution of} frequencies.\\
     {The distribution of frequencies shows some dispersion. Models, therefore, need to be able 
    to explain the origin of this distribution and on what it depends.}\\
   
     \item The model should be able to explain the HFQPO frequencies' selection mechanism:  \\
     We have seen that HFQPOs appear in pairs or alone.  Models, therefore, need to be able 
    to explain the selection mechanism that favours the occurrence of one of the HFQPOs instead of the other, and the appearance of 
    pairs in some cases and single peaks in others.  This is particularly obvious from the comparison of group IV and V that show 
    relatively similar continuums but different QPO complexes at high frequency.  \\

     \item an HFQPO model should be able to exist in a wide variety of initial conditions.\\
	Indeed, strong LFQPOs are often observed in the presence of 
   HFQPOs,  therefore it is required to have a regime where both models exist simultaneously in the system. This is especially important as 
   LFQPOs tend to have a much stronger influence on the system and will change the state/initial condition in which the HFQPO model is developing. 
 \end{itemize}

\section{Summary}

	We created five groups of combined PDSs by adding together observations having similar HFQPOs, continuums, and LFQPOs  from the  1998--99 and 2000 outbursts of XTE J$1550$-$564$. The HFQPO frequencies span from $140$ to $283$Hz respectively from group I to V. The improved statistics of the combined PDSs leads to two probable 
	detections of a pair of HFQPOs in group III and group V.\\
	The most striking feature of those pairs is that their frequency ratios are not in the previously detected 
	$3$:$2$ ratio even though we
	used some of the observations in which they were previously detected. Instead the frequency ratio in these two cases 
	is closer to 0.75. If one assumes that the two peaks are harmonically related, then the features 
	would need to be the third and 
	fourth harmonics of a yet to be detected fundamental at a much lower frequency. Because we only added together 
	observations having similar originally detected frequencies and  	
	continuum, this means that the $3$:$2$ ratio is not always present in XTE J$1550$-$564$. In turn, theoretical models need to
	 focus on explaining the frequency
	selection mechanism which gives two  $3$:$4$ HFQPOs in some cases, and also why a pair is present in group V and  not in
	 group IV, while the two PDSs do show very similar behavior.

\begin{acknowledgements}
The authors thank the anonymous referee that helped to clarify several points an to improve the paper to this final form.
We acknowledge the financial support of the UnivEarthS Labex program at Sorbonne Paris Cite (ANR-10-LABX-0023 and ANR-11-IDEX-0005-02).
JR acknowledges  funding support from the French Research National Agency: CHAOS project ANR-12-BS05-0009 (\texttt{http://www.chaos-project.fr}).
\end{acknowledgements}

\bibliography{HFQPO_ApJ_V3}
\end{document}